# Interface Engineering Enabled Low Temperature Growth of Magnetic Insulator on Topological Insulator

*Nirjhar Bhattacharjee, Krishnamurthy Mahalingam, Alexandria Will-Cole, Yuyi Wei, Adrian Fedorko, Cynthia T. Bowers, Michael Page, Michael McConney, Don Heiman, Nian Xiang Sun\**

N. Bhattacharjee 1, A. Will-Cole 1, Yuyi Wei 1, Prof. N. X. Sun 1
1 Northeastern University, Department of Electrical and Computer Engineering, Boston MA 02115, USA
E-mail: n.sun@northeastern.edu (Corresponding author)

A. Fedorko 2, Prof. D. Heiman 2
2 Northeastern University, Department of Physics, Boston MA 02115, USA

Prof. D. Heiman 3,
3 Plasma Science and Fusion Center, MIT, Cambridge, MA 02139, USA

Dr. K. Mahalingam 4, C.T. Bowers 4, Dr. M. Page 4, Dr. M. McConney 4
4 Air Force Research Laboratory, Nano-electronic Materials Branch, Wright Patterson Air Force Base, OH 05433, USA
USA

## Abstract

Combining topological insulators (TIs) and magnetic materials in heterostructures is crucial for advancing spin-based electronics. Magnetic insulators (MIs) can be deposited on TIs using the spin-spray process, which is a unique non-vacuum, low-temperature growth process. TIs have highly reactive surfaces that oxidize upon exposure to atmosphere, making it challenging to grow spin-spray ferrites on TIs. In this work, it is demonstrated that a thin titanium capping layer on TI, followed by oxidation in atmosphere to produce a thin $TiO_x$ interfacial layer, protects the TI surface, without significantly compromising spin transport from the magnetic material across the $TiO_x$ to the TI surface states. First, it was demonstrated that in $Bi_2Te_3/TiO_x/Ni_{80}Fe_{20}$ heterostructures that $TiO_x$ provided an excellent barrier against diffusion of magnetic species, yet maintained a large spin-pumping effect. Second, the $TiO_x$ was also used as a protective capping layer on $Bi_2Te_3$, followed by the spin-spray growth of the MI, $Ni_xZn_yFe_2O_4$ (NZFO). For the thinnest $TiO_x$ barriers, $Bi_2Te_3$/$TiO_x$/NZFO samples had AFM disordered interfacial layer because of diffusion. With increasing $TiO_x$ barrier thickness, the diffusion was reduced, but still maintained strong interfacial spin-pumping interaction. These experimental results demonstrate a novel method of low-temperature growth of magnetic insulators on TIs enabled by interface engineering.

1. Introduction

Realization of the quantum anomalous Hall (QAH) effect has been demonstrated in several magnetic topological insulators (MTIs). MTIs can be synthesized by doping magnetic elements into topological insulators (TIs) [1-9] or growth of intrinsic MTI compounds [10-19]. However, these MTI materials require temperatures much lower than their magnetic $T_c$ to achieve quantization in the Hall effect. This deficiency has been attributed to inhomogeneity in magnetic doping which causes localized variation in surface state exchange gaps. Magnetism can also be induced in the surface of TI thin films via proximity exchange effect. This proximity induced magnetization (PIM) is envisioned by coupling TIs with a magnetic insulator (MI). PIM in TIs have been experimentally confirmed in heterostructures of TIs with magnetic insulators (MIs) such as rare earth garnets and EuS [20-28], MTIs [29,30] and telluride van der Waals (vdW) magnets [31]. However, recent studies have pointed out the complexities in measuring proximity exchange in these systems, [31] which can be affected by a number of problems such as diffusion at the interface and band bending. Thus, providing a barrier for diffusion while permitting magnetic exchange in TI/MI heterostructures will be necessary for devices, and it is shown here that a $TiO_x$ interface barrier achieves this goal. Only a single study by Watanabe et al. [31] has shown PIM and QAH in a TI/MI/TI thin film sandwich structure of (Zn,Cr)Te/$(Bi,Sb)_2Te_3$/(Zn,Cr)Te. However, even in this material system, the QAH effect was seen at extremely low temperature of 0.03 K. This low temperature is possibly due to: (1) High-quality MIs which were grown on TI thin films have very low Curie temperature, $T_c$, such as (Zn,Cr)Te with $T_c < 40$ K [31] and EuS with $T_c$ of ~17 K [25]. (2) Even with 20-25% of the magnetic species, Cr in (Zn,Cr)Te finite localized non-magnetic nano-regions may exist in.

MIs such as rare-earth garnets and ferrites are excellent materials that are magnetically ordered well above room-temperature, making them ideal candidates for QAH materials coupled with TIs. However, growth of MIs on top of TIs using a high vacuum process is challenging because of their high growth temperatures (> 600 °C) [32]. At these temperatures, TIs are thermally unstable [33,34] and can melt or sublimate making them unusable for TI/MI QAH systems. The spin-spray process provides a solution for growth of MIs at much lower temperatures of ~100 °C [35-39]. Ferrites are MIs with a high $T_c$ (~750 °C), which makes them excellent candidates for realization of QAH and spintronic device applications, provided they can be heterostructured with TIs. However, an engineering challenge remains to be solved. TIs have a highly reactive surface, which undergoes reactions promoted by the topological surface states (TSS) electrons [40-43]. Unfortunately, exposure to atmosphere leads to rapid oxidation of Bi, Sb, Se and Te, which is experimentally observed in this work, decouples the MI layer from the adjacent TI. Without a finite exchange interaction between the TSS and the magnetic material, none of the spintronic and quantum properties can be achieved.

In this work, interface engineering of TI/FM and TI/MI heterostructures using an ultra-thin $TiO_x$ interlayer will be presented. The $TiO_x$ interlayer has three important properties: (1) it protects the TI surface against oxidation when exposed to atmosphere after the TI growth; (2) it limits interdiffusion at the TI/FM and TI/MI interface; and (3) it maintains sufficient transparency for spin transport that is crucial for devices. The $TiO_x$ insertion layer acts as a barrier to interfacial diffusion of elements without significantly compromising interaction of magnetic moments with TSSs. First, the interfacial diffusion was examined with and without $TiO_x$ on samples of $Bi_2Te_3/Ni_{80}Fe_{20}$ and spin-spray grown NZFO. Second, the spin transparency of these heterostructures was examined by measuring the spin-orbit torque (SOT) effects on the Gilbert damping in FMR experiments. Further information was obtained from magnetization measurements, TEM images and EDS scans. From these studies it is evident that low-temperature spin-spray growth of MIs on TIs can be achieved by interface engineering of TIs which is a significant achievement. This is expected to lead to further experimental explorations of interfacial topological properties on the surface of TI/FM, TI/MI, MI/TI and MI/TI/MI heterostructures and their eventual integration of these materials systems into practical quantum and spintronic devices.

## 2. Experimental Results and Discussion

The experimental section is organized as follows: (2.1) contains results on the effects of TI surface oxidation and interlayer $TiO_x$ on spin-pumping and interdiffusion; (2.2) describes the formation of an AFM interlayer via atomic diffusion; (2.3) contains results on TI/MI heterostructures grown by non-vacuum, low-temperature spin-spray deposition of NZFO on $Bi_2Te_3$.

## 2.1. Effects of TI Surface Oxidation an Interlayer $TiO_x$ on Spin-pumping and Interfacial Diffusion

### 2.1.1. Suppression of SOT on TI Surface due to Surface Oxidation

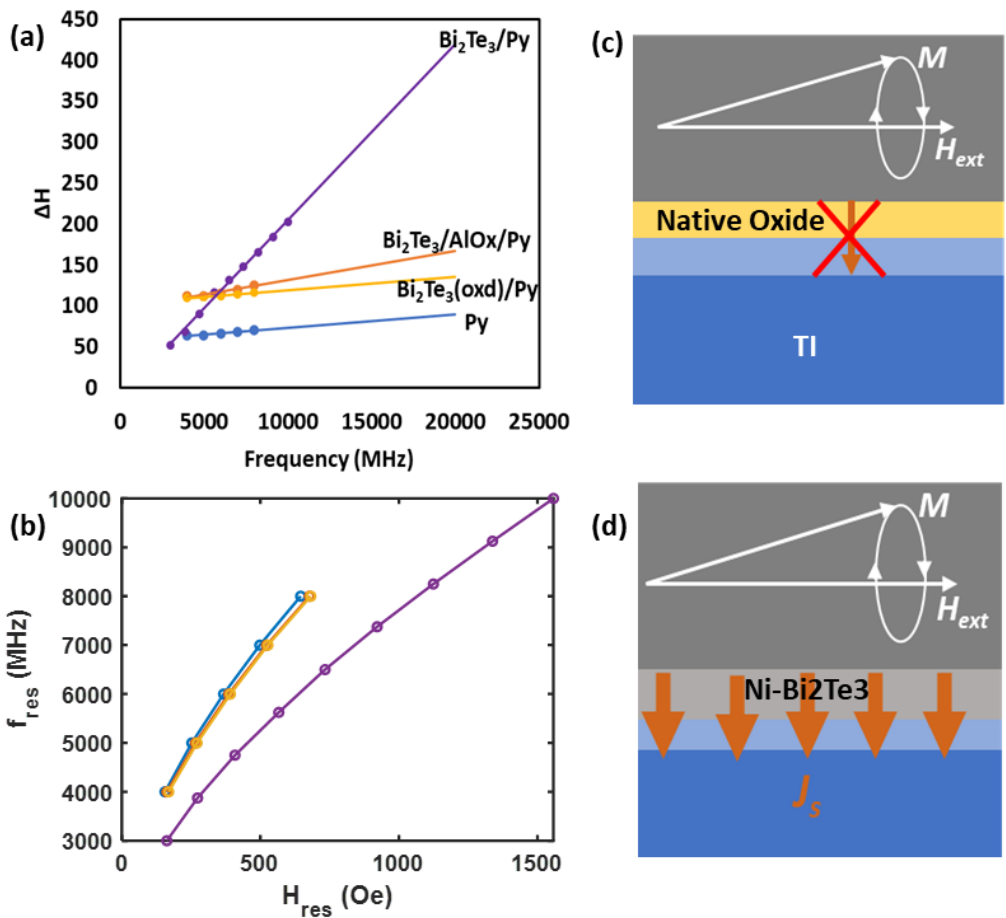

*Figure 1. Results of FMR measurements on TI/FM heterostructures with differing interfacial conditions. a) FMR linewidth ΔH as a function of FMR frequency for $Bi_2Te_3$/Py, $Bi_2Te_3$ (oxidized in atmosphere)/Py and Py thin films, and b) FMR frequency as function of FMR resonant field and*

*Kittel equation fitting for Py, Bi$_2$Te$_3$/AlO$_x$/Py, Bi$_2$Te$_3$ (oxidized)/Py and Bi$_2$Te$_3$/Py samples. c) Illustration of suppressed spin pumping due to oxidation of Bi$_2$Te$_3$ on exposure to atmosphere, and d) large spin pumping in Bi$_2$Te$_3$/Py with Ni-Bi$_2$Te$_3$ interfacial layer resulting from diffusion of Ni [42].*

The spin transparency of TI/FM interfaces is conveniently determined in FMR experiments by measuring changes in the Gilbert damping. In high-quality TI/FM interfaces the precessing magnetization in the FM results in spin-pumping into the TI. This is because the strong SOC in TIs, resulting in ISHE and enhanced gilbert damping. On the other hand, no spin-pumping is seen when the spins are prohibited from transferring from the FM to the TI layer. On exposure to atmosphere, TI surfaces readily oxidize [40], which leads to topologically trivial insulators on the surface. Almost total loss of spin-pumping was observed in TI/FM heterostructures when the Bi$_2$Te$_3$ thin film was exposed to the atmosphere before deposition of the FM (20 nm thick Py).

To understand the effect of surface oxidation of TI on spin-pumping in TI/FM heterostructures, FMR experiments were performed at room temperature with the following samples: Bi$_2$Te$_3$/Py, Bi$_2$Te$_3$(oxidized in atmosphere)/Py, Bi$_2$Te$_3$/AlO$_x$ (3nm Al oxidized in atmosphere)/Py, and Al(10nm)/Py (control). For the sample Bi$_2$Te$_3$(oxidized)/Py as the label suggests, Bi$_2$Te$_3$ was exposed to atmosphere before deposition of Py. The FMR linewidths $\Delta H$ were extracted by fitting a Lorentzian function to the FMR field scans (see Experimental Methods section and Supporting Information Section S1). The Gilbert damping $\alpha$ was extracted from a linear fit to the $\Delta H$ as a function of frequency using the relation, $\Delta H = \Delta H_0 + \frac{2\pi}{\gamma}\alpha$, as shown in Figure 1a. Here, $\Delta H_0$ is the inhomogeneous linewidth broadening and $\gamma$ is the gyromagnetic ratio ($\frac{\gamma}{2\pi} = 2.8$ MHz/Oe for metallic films) [48]. Figure 1a shows that the Bi$_2$Te$_3$/Py sample has a very large Gilbert damping, $\alpha = 0.060$, compared to all the other samples, indicating that spin transport is nearly eliminated by adding interfacial layers. The sample with $Bi_2Te_3$(oxidized)/Py had a much reduced $\alpha = 0.004$, almost the same as a control sample of Al/Py. Note that when the metal-oxide is thick, as with the $Bi_2Te_3/AlO_x$/Py sample, $\alpha = 0.0036$, which is higher than the control sample but still much lower than that of the $Bi_2Te_3$/Py. However, note that in the $Bi_2Te_3$/Py sample there is a naturally-diffused interlayer of antiferromagnetic (AFM) Ni-$Bi_2Te_3$ that remains effective for spin transfer and large damping [42].

Further, the effective magnetization, $4\pi M_{eff}$ was extracted by fitting the FMR frequency, $f_{res}$ as function of FMR, $H_{res}$ field plot using the Kittel equation, $f_{res} = \frac{\gamma}{2\pi}\sqrt{(H_{res} - H_a)(H_{res} - H_a + 4\pi M_{eff})}$ as shown in Figure 1b. Here, $H_{res}$ is the FMR resonance field and $H_a$ is the uniaxial anisotropy field. A significant change in $4\pi M_{eff}$ suggests change in magnetic anisotropy because of interfacial exchange interaction of the FM layer with TSS [46]. Further, exchange interaction of an FM with a large SOC material should also result in change in $\gamma$. However, due to the large thickness of the FM layer, this change is expected to be negligibly small [48]. Hence, a $\frac{\gamma}{2\pi}$ value of 2.8 MHz/Oe has been considered for the metallic FMs in this work. The fitting to Kittel equation reveals $4\pi M_{eff}$ values of 1.22 kOe for Py, slightly reduced values of 1.15 kOe for $Bi_2Te_3$ (oxd)/Py, 1.14 kOe for $Bi_2Te_3/AlO_x$/Py and a dramatically reduced value of 6.55 kOe for the $Bi_2Te_3$/Py heterostructure. This suggests reduced interfacial magnetic interactions in in the $Bi_2Te_3$(oxidized)/Py and $Bi_2Te_3$/AlOx/Py samples because of decoupling of the FM from the TSS of the TI represented by the schematics in Figures 1c,d. In MBE grown TI thin film samples, surfaces are usually capped with Se or Te [49-55], which is evaporated by heating in a high vacuum before growing the FM layer in SOT devices. The presence of a high-quality capping layer becomes especially significant for successfully growing MI ferrites using the low-temperature spin-spray process on TIs as substrates are exposed to the atmosphere during deposition. The choice of an effective capping/interface engineered layer is essential as interfacial magnetic interactions and spin transparency must not be significantly diminished through of the barrier for efficient spintronic device applications.

### 2.1.2. TiO$_x$ Interlayer Effects on Diffusion in TI/FM Heterostructures

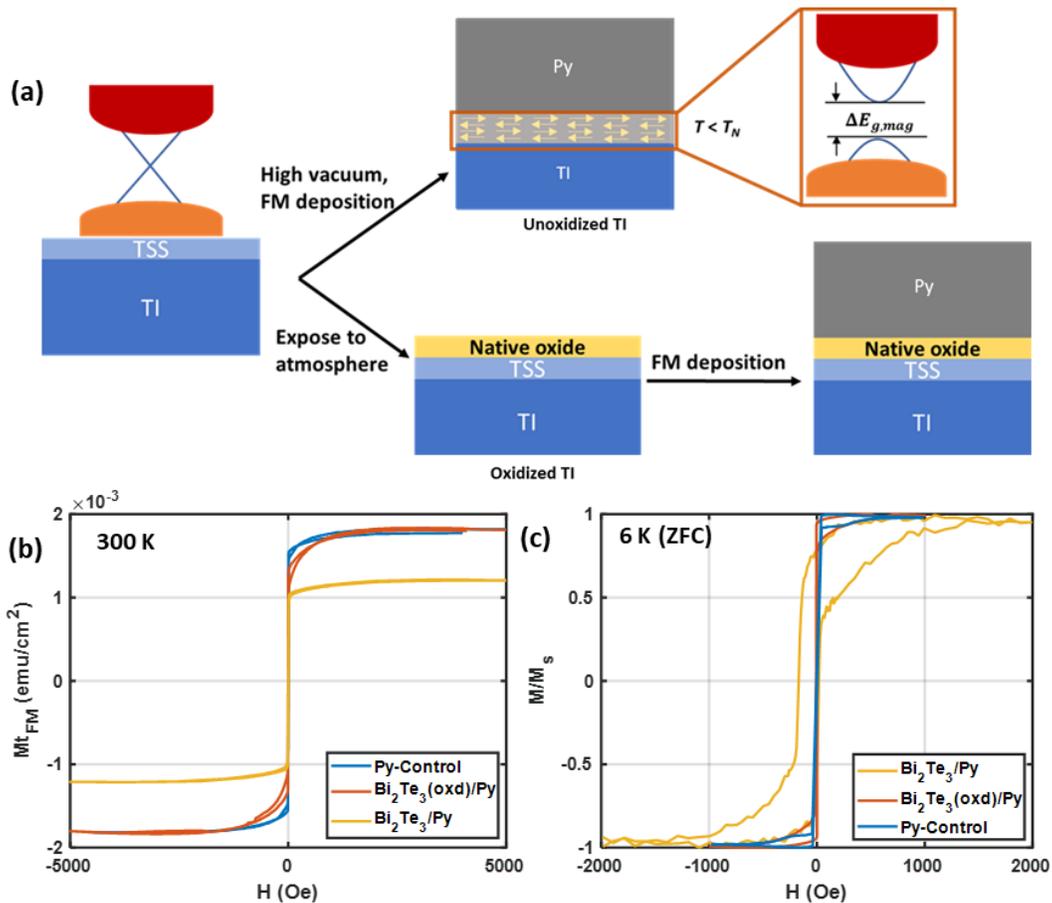

*Figure 2. Results of TI/FM heterostructure with and without oxidized TI. a) Schematic representation of effect of oxidation of Bi$_2$Te$_3$ before deposition of FM Py. Deposition of Py on*

*Bi$_2$Te$_3$ without breaking vacuum results in interfacial AFM interface which hypothetically can lead to magnetic gap opening in TSS. Bi$_2$Te$_3$ on exposure to atmosphere oxidizes the surface and pushes the TSS below the oxidized surface and the magnetic Py layer becomes decoupled from the Bi$_2$Te$_3$ surface. b) Reduction in magnetic moments arise from diffusion of Ni from Py and reacting with Bi$_2$Te$_3$. Oxidized Bi$_2$Te$_3$ does not show diffusion and moment reduction effects. c) Large exchange bias of 82 Oe in Bi$_2$Te$_3$/Py generated by AFM NiBi$_2$Te$_4$ in the interface [42].*

The atomic interdiffusion in TI/FM samples was investigated as a function of TiO$_x$ interlayers. Two cases are illustrated in Figure 2a, one where the Py is grown on the TI without breaking the chamber vacuum and the other where the TI was exposed to atmosphere before depositing the Py. Without breaking the vacuum, the Bi$_2$Te$_3$/Py interface develops a topological nontrivial AFM interfacial layer because of diffusion and reaction of Ni with Bi$_2$Te$_3$ and forming NiBi$_2$Te$_4$[42,43]. In contrast, oxidizing the surface of the TI prevents reaction and diffusion of Ni or Fe from Py and should not allow for forming a topological interfacial layer. To study the effect of surface oxidation of Bi$_2$Te$_3$ on the interface diffusion and reactions in Bi$_2$Te$_3$/Py, room temperature $m(H)$ measurements were performed on the samples of Bi$_2$Te$_3$/Py, Bi$_2$Te$_3$(oxidized)/Py and Py (control) samples. Because of interfacial diffusion of Ni, Figure 2b shows that the saturation moment of the Py layer decreases by ~40%. This happens because the diffused Ni reacts with Bi$_2$Te$_3$ to form compounds which are not ferromagnetically ordered at room temperature [42]. But as expected, no magnetic moment is lost when the Bi$_2$Te$_3$ is oxidized before deposition of the Py layer, as the oxide interlayer prevents diffusion of Ni out of the Py layer. Furthermore, as a result of inhibiting diffusion, the oxide interlayer prohibits Ni from reacting with the Bi$_2$Te$_3$, which prevents formation of an AFM NiBi$_2$Te$_4$ interlayer. The presence of an AFM interfacial layer next to the FM Py was found to induce a large spontaneous exchange bias field, $H_{EB}$ ~ 80 Oe [42] at low temperatures (ZFC 6 K). However, as shown in Figure 2c, samples of $Bi_2Te_3$(oxidized)/Py didn't show any exchange-bias effects in contrast to the large exchange bias in $Bi_2Te_3$/Py heterostructures. These results confirm the suppression of surface reactivity due to atmospheric oxidation of TIs. The oxides formed on the TI surface decouple the TSS chemically from the adjacent FM layer, although we will show that spin transfer can remain effective. These results further emphasize the need for interface engineering for integration of TIs with spin-spray grown MIs.

### 2.1.3. Interface Engineering in TI/FM Heterostructures using $TiO_x$

It is important to investigate how the thickness, $z_{TiOx}$ of interlayers of Ti and its oxide $TiO_x$ in TI/FM structures affect the SOT and related magnetic properties. Ti possesses negligible SOC and is highly transparent to spin currents [45], which makes $TiO_x$ an excellent material for interface engineering of TIs with FM oxides for spintronic device applications. In order to understand the effects of the $TiO_x$ insertion layers in TI/FM heterostructures, crystalline *c*-axis-oriented $Bi_2Te_3$ was grown using magnetron sputtering similar to the work in reference [42]. The growth of high-quality $Bi_2Te_3$ was followed by deposition of ultrathin Ti that was subsequently exposed to atmosphere. Py thin films of thickness 20 nm were then grown on the $Bi_2Te_3/TiO_x$ bilayers to first study their effect on magnetic properties. Figure 3a shows the effects of interfacial diffusion and solid-state reactions by monitoring the reduction of Py moments. Ni and small amounts of Fe diffuse out from the Py and form interfacial compounds due to solid state reactions with $Bi_2Te_3$ [42,43]. Looking at the saturation moment, it is clear that $TiO_x$ layers with thickness ≤ 1 nm cannot prevent interfacial diffusion of the magnetic species, which is aided by surface reaction with Bi$_2$Te$_3$. Note that the cited TiO$_x$ thickness, $z_{TiOx}$ is that of the Ti thin film deposited on a Si/SiO$_2$ substrate. The diffusion of primarily Ni and small amounts of Fe, leads to a moment loss of ~40%, which is comparable to the control Bi$_2$Te$_3$/Py sample [42]. However, TiO$_x$ insertion layers of thickness ≥ 2 nm act as excellent barriers to the diffusion of Ni from Py preventing any loss of magnetic moments.

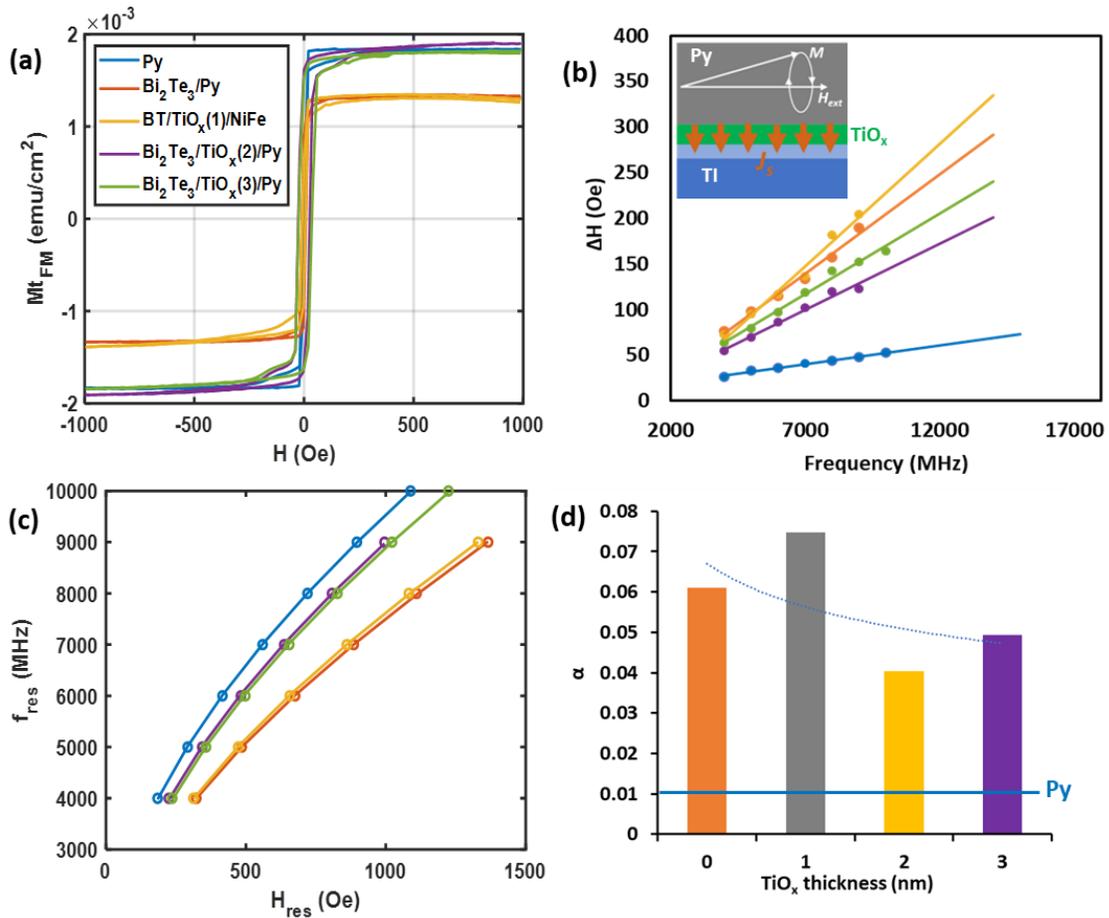

*Figure 3. Results of magnetization and FMR experiments on Bi$_2$Te$_3$/Py structures with TiO$_x$ interlayers. a) m(H) loops measured at room temperature, showing no loss of moment for the two thicker TiO$_x$ layers. b) FMR linewidth ΔH as a function of FMR frequency, showing strong damping for all thickness of TiO$_x$ interlayers. Inset illustrates the spin current passing down thru the TiO$_x$. c) FMR frequency as function of FMR field and Kittel equation fitting. d) Summary of*

*Gilbert damping for Py, and for $z_{TiOx}$ = 0, 1, 2, 3 nm thick TiO$_x$ interlayers, which far exceeds that of the Py control sample.*

Next, spin-pumping effect was investigated as a function of TiO$_x$ thickness via the enhancement in Gilbert damping in FMR experiments. A dramatic improvement in spin pumping over AlO$_x$ insertion layers was observed, which can be seen by comparing the results of Figure 1a with those of Figure 3b-d. Table 1 reports the values of various properties for the Bi$_2$Te$_3$/TiO$_x$/Py samples. Values of $\alpha$ obtained for the Bi$_2$Te$_3$/TiO$_x$/Py samples were 0.074, 0.040 and 0.049 for the 1, 2 and 3 nm thick TiO$_x$ layers, respectively, and are comparable to that of 0.061 for the Bi$_2$Te$_3$/Py sample. All these samples hence show significantly enhanced $\alpha$ compared to 0.011 for the control sample of Py, thus confirming a large spin-pumping effect. The spin-mixing conductance can be quantified using the spin-mixing conductance parameter given by $g_{\uparrow\downarrow} = 4\pi M_s t_{FM} \frac{\Delta\alpha}{\hbar\gamma}$, where $t_{FM}$ is the thickness of the FM layer, $\Delta\alpha$ is the enhancement in $\alpha$, $\hbar$ is the reduced Plank's constant and $\gamma$ is the gyromagnetic ratio. The enhancement in $\alpha$ for the Bi$_2$Te$_3$/Py and Bi$_2$Te$_3$/TiO$_x$(1nm)/Py samples may also contain some contributions from the loss of magnetic moments due to the large interdiffusion. But, because of the complexity of the interfaces of these samples, the effect of spin-pumping may not be accurately isolated. The $g_{\uparrow\downarrow}$ values reported for these samples were calculated based on the assumption that the enhancement in $\alpha$ compared to control sample of Py is due to spin-pumping only, which may nevertheless be incorrect. However, as the Bi$_2$Te$_3$/TiO$_x$/Py samples with 2 nm and 3 nm TiO$_x$ barriers had no noticeable interfacial diffusion of Ni, Fe, they had large $g_{\uparrow\downarrow}$ values as listed in Table 1. These large values indicate that TIs with TiO$_x$ interface-engineered heterostructures can be excellent candidates for efficient SOT devices. Further, Table 1 also lists saturation magnetization, $4\pi M_s * t_{FM}$ determined from the *m(H)* loop measurements in Figure 3a. Using those values, the effective interfacial out-of-plane (OOP) anisotropy field values can be calculated using the relation, $H_{OOP} = 4\pi M_s - 4\pi M_{eff}$. The $H_{OOP}$ values listed in Table 1 show that Bi$_2$Te$_3$/TiO$_x$ (2 and 3 nm)/Py samples have $H_{OOP}$ = 2.02 and 2.15 kOe, respectively. The negative $H_{OOP}$ for the Py sample indicates that the magnetic anisotropy and easy axis of the sample is highly in-plane as expected. The Bi$_2$Te$_3$/TiO$_x$/Py samples on the other hand show a large OOP anisotropy because of the exchange interaction of the interfacial magnetic moments of Py with the TSS of Bi$_2$Te$_3$.

**Table 1. Properties for Bi$_2$Te$_3$/TiO$_x$/Py heterostructures for different TiO$_x$ interlayer thickness. The parameters are described in the text.**

| Material | $z_{TiOx}$ (nm) | $M_s*t_{FM}$ (emu/cm$^2$) | $\alpha$ | $g_{\uparrow\downarrow}$ (m$^{-1}$) | $4\pi M_{eff}$ (kOe) | $H_{OOP}$ (kOe) |
|---|---|---|---|---|---|---|
| Py | - | $1.83 \times 10^{-3}$ | 0.011 | - | 12.6 | -1.05 |
| Bi$_2$Te$_3$/Py | 0 | $1.33 \times 10^{-3}$ | 0.061 | $1.70 \times 10^{20}$ | 6.32 | - |
| Bi$_2$Te$_3$/TiO$_x$/Py | 1 | $1.28 \times 10^{-3}$ | 0.074 | $2.17 \times 10^{20}$ | 6.56 | - |
| Bi$_2$Te$_3$/TiO$_x$/Py | 2 | $1.82 \times 10^{-3}$ | 0.040 | $9.88 \times 10^{19}$ | 9.53 | 2.02 |
| Bi$_2$Te$_3$/TiO$_x$/Py | 3 | $1.75 \times 10^{-3}$ | 0.049 | $1.29 \times 10^{20}$ | 9.40 | 2.15 |

## 2.2. Tuning of Topological AFM Interface with TiO$_x$ Barrier

Formation of an interfacial topological AFM layer generated by diffusion of Ni and solid-state reaction with Bi$_2$Te$_3$ was shown previously in ref [42,43]. The solid-state reaction was found to result in loss of magnetic moment and emergence of a large spontaneous exchange bias in Bi$_2$Te$_3$/Py heterostructures. The $m(H)$ loop measurements at room temperature in Figure 4a show loss of magnetic moments in the Bi$_2$Te$_3$/TiO$_x$/Py sample with the 1 nm thick TiO$_x$, similar to the Bi$_2$Te$_3$/Py sample deposited with the same growth conditions. The increase in TiO$_x$ barrier thickness $z_{TiOx}$ was found to prevent the diffusion of Ni across the interface. In order to understand the effect of the interface in the Bi$_2$Te$_3$/TiO$_x$/Py samples, $m(H)$ loops were measured at 300 and 6

K in zero-field cooling (ZFC) conditions. The sample with $z_{TiOx} \approx 1$ nm clearly shows a large $H_{EB}$ = 56 Oe and an enhanced $H_c$ = 66 Oe, as shown in Figure 4a. This arises from exchange interaction between the FM Py layer with the interfacial AFM layer [42]. This $H_{EB}$ was however smaller than the $H_{EB}$ = 82 Oe for $Bi_2Te_3$/Py without a $TiO_x$ barrier. This is due to the reduction in exchange interaction caused by the insertion of the ultra-thin $TiO_x$ layer between the interfacial AFM Ni-$Bi_2Te_3$ and the Py layer. With an increase in thickness of the $TiO_x$ barrier to $z_{TiOx} \geq 2$ nm, there is no diffusion of Ni and hence no $H_{EB}$ is observed in the at low temperatures. This clearly shows the effect of interface engineering on controlling the interfacial properties in TI/FM heterostructures that will be essential for spintronic devices. These results also show the potential of exploring fascinating interfacial phases in TI/FM heterostructures.

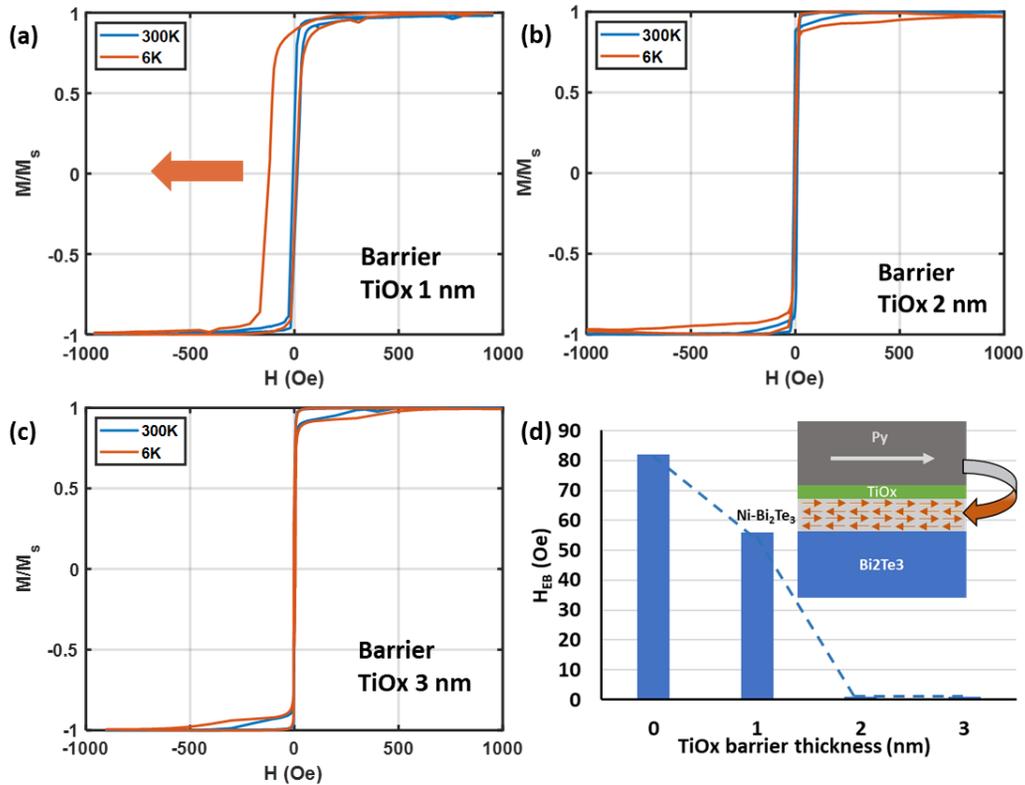

*Figure 4. Measurements of m(H) loop at 6 K (ZFC) and 300 K showing exchange bias of Bi$_2$Te$_3$/TiO$_x$/Py samples. a) H$_{EB}$ = 56 Oe for z$_{TiOx}$ = 1 nm at 6 K. b) No exchange bias for z$_{TiOx}$ = ~2 nm, and c) no exchange bias for z$_{TiOx}$ = ~3 nm. d) Comparison of exchange bias at 6 K for the samples. (Inset shows the diffusion-generated AFM layer below the TiO$_x$ for z$_{TiOx}$ = ~1 nm.*

## 2.3. Bi$_2$Te$_3$/TiO$_x$/NZFO Heterostructures

### 2.3.1. Interfacial Morphology of Bi$_2$Te$_3$/TiO$_x$/NZFO Heterostructures

The following describes results of non-vacuum, spin-spray growth of the magnetic insulator Ni$_x$Zn$_y$Fe$_2$O$_4$ on Bi$_2$Te$_3$, which is facilitated using a TiO$_x$ interlayer. Bi$_2$Te$_3$/TiO$_x$/NZFO heterostructures were investigated for their morphology, diffusion and spin-pumping. The morphology of the Bi$_2$Te$_3$/TiO$_x$/NZFO interface was first studied using cross-section TEM imaging and EDS. As shown in Figures 6a, the Bi$_2$Te$_3$ is vdW-layered and highly *c*-axis oriented as expected [42]. The TEM image also shows a disordered interfacial layer for $z_{TiOx}$ = 1 nm that is created by diffusion and solid-state reactions. In contrast, as observed in the TEM images in Figures 6b-c, with increase in thickness of the TiO$_x$ barrier layer to 2 nm and 3 nm the thickness of the disordered interface is significantly reduced progressively. The EDS line profiles in Figures 6d-f show the intensity of the EDS signal. With the 1 nm TiO$_x$ barrier layer there is a clear interfacial layer primarily consisting of Bi, Te, Fe and O (Figure 5.6a) with O significantly diffusing ~10 nm into Bi$_2$Te$_3$. This is possibly the primary factor for creation of the highly disordered interfacial layer observed in Figure 6a. For the 2 and 3 nm thick TiO$_x$ interlayer the thickness of the oxygen diffusion is reduced to ~5 nm and ~0 nm, respectively. Further, thickness of TiO$_x$ > 2nm prevents any Fe diffusion as confirmed by TEM images as well as EDS line profiles. This demonstrates the effectiveness of the thicker TiO$_x$ barrier layers in the Bi$_2$Te$_3$/TiO$_x$/NZFO heterostructures.

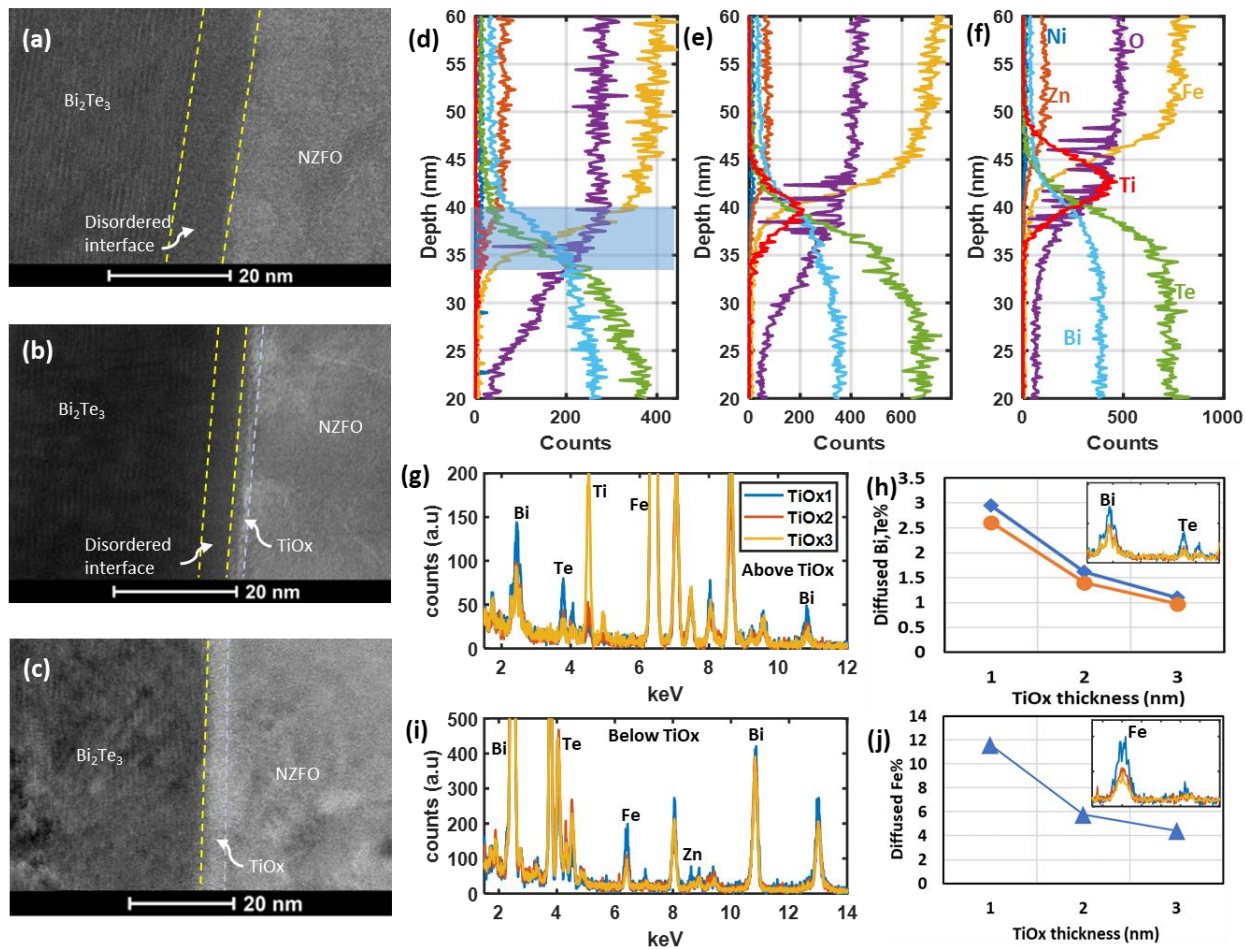

*Figure 5. TEM images for $Bi_2Te_3/TiO_x/NZFO$ with $TiO_x$ thickness: a) 1 nm b) 2 nm, and c) 3 nm, showing c-axis-oriented vdW layered structure and interfacial layers. EDS line profiles of the TEM images for: d) 1 nm, e) 2 nm, and f) 3 nm samples. g) Comparison of EDS spectra above $TiO_x$ barrier. Concentration of diffused Bi and Te: h) above $TiO_x$ barrier and i) below TiOx barrier for comparison. j) Diffused Fe below $TiO_x$ barrier in the 1, 2, and 3 nm samples.*

Diffusion of elements at the interface was analyzed with EDS by extracting the relative concentration of Bi and Te above the $TiO_x$ and Fe below the $TiO_x$ regions. The concentration of oxygen couldn't be analyzed accurately as EDS results for light elements are inaccurate. For this analysis of interfacial diffusion only a small thickness of about 5 - 6 nm above and below the $TiO_x$ regions were considered (see Supporting Information Section S4). As shown in Figure 6h, which is extracted from the Figure 6g, there is a small diffusion of Bi and Te across the interface to the NZFO side. Similarly, Figure 6j, which is extracted from the EDS spectra in Figure 6i, also shows diffusion of Fe from NZFO into $Bi_2Te_3$. The diffused Bi relative atomic% values are 2.9%, 1.6% and 1.1% for the samples with 1, 2 and 3 nm $TiO_x$, respectively. The values for diffused Te also show a similar trend of 2.6%, 1.4% and 1.0%, respectively. Interestingly, the diffusion of Fe shown in Figure 6j below the $TiO_x$ barrier layers from NZFO to the $Bi_2Te_3$ side was much higher at 11.6% for the sample with 1 nm $TiO_x$ and reduces significantly to 5.7% and 4.4% for the samples with 2 and 3 nm $TiO_x$ barriers, respectively. The concentrations of Ni and Zn are very small ($< 2\%$) throughout the cross-section of the NZFO layer and are within the noise range of EDS signal as shown in Figures 6g,i. Hence, diffusion of these elements could not be determined accurately. Nevertheless, the significant diffusion of Fe in the 1 nm $TiO_x$ samples form a disordered interface that results in a fascinating AFM interfacial phase which will be described in Section 2.3.3.

### 2.3.2. Room Temperature Magnetic Properties of $Bi_2Te_3/TiO_x/NZFO$ Heterostructures

NZFO thin films of thickness ~250 nm were grown on $Bi_2Te_3/TiO_x$ bilayers using the method described in refs. [38,39] (see Experimental Methods). The deposition was made at a substrate temperature of ~100°C, which is not expected to cause any phase change in the $Bi_2Te_3$ layer. The Gilbert damping $\alpha$ extracted from the slope of the FMR linewidth versus frequency in Figure 6a clearly shows a smooth decreasing trend with increasing $TiO_x$ thickness, shown in Figure 6b and listed in Table 2. The NZFO control sample had an $\alpha$ value similar to the values of the $Bi_2Te_3/TiO_x/Py$ samples with 2 nm and 3 nm $TiO_x$. The significantly enhanced $\alpha$ for the 1 nm $TiO_x$ insertion layer indicates a large spin-pumping and possibly some spin memory-loss (SML) effects in the $Bi_2Te_3/TiO_x/NZFO$ heterostructure [46,56]. Because of the large thickness of the NZFO layer, the overall Gilbert damping enhancement effect is reduced in the $Bi_2Te_3/TiO_x/NZFO$ with $z_{TiOx}$ = 2 and 3 nm samples. This is due to the increase in distance between the MI and TI because of the presence of the TiOx insertion layer. The enhancement in Gilbert damping indicates that the diffused oxygen in the samples is extremely low and does not significantly affect the interfacial spin transparency, unlike TIs exposed to atmosphere. The decreasing trend in $\alpha$ with the increase in $z_{TiOx}$ is similar to the trend observed with Py as the FM layer. In contrast to thick metallic FM films, MIs show more pronounced changes in magnetism because of proximity to large SOC materials [46,48]. Fitting the Kittel equation to the $f_{res}$ versus $H_{res}$ relation, the values of $\frac{\gamma}{2\pi}$ and $4\pi M_{eff}$ of the samples as shown in Figure 7c and Table 2. The $4\pi M_{eff}$ values for $z_{TiOx}$ = 2 and 3 nm were very close to the NZFO control sample. However, the $z_{TiOx}$ = 1 nm sample showed a dramatic reduction to less than half the other values. This change in $4\pi M_{eff}$ has possible contributions from: (1) the exchange interaction between the magnetic moments in the MI and TI, as well as (2) new materials phases formed from diffusion and solid-state reactions in the interface, and (3) PIM in the TI due to the adjacent MI. However, these effects could not be isolated in this study and will need further analysis in the future. Furthermore, the $\frac{\gamma}{2\pi}$ for the Bi$_2$Te$_3$/TiO$_x$/NZFO samples obtained from the Kittel equation fittings in Figure 6c and listed in Table 2 were significantly different larger compared to the NZFO control sample. The large changes in $\gamma$ plotted in Figure 6d signify strong exchange interaction effect of the magnetic moments in the MI with the TSSs in the TI surface [46]. The HRTEM and EDS measurements in the earlier section clearly indicated lack of diffusion in the Bi$_2$Te$_3$/TiO$_x$/NZFO (and Py) when TiO$_x$ has a thickness larger than 2 nm. This confirms a finite exchange interaction between the TSS and MI even with a finite separation between them.

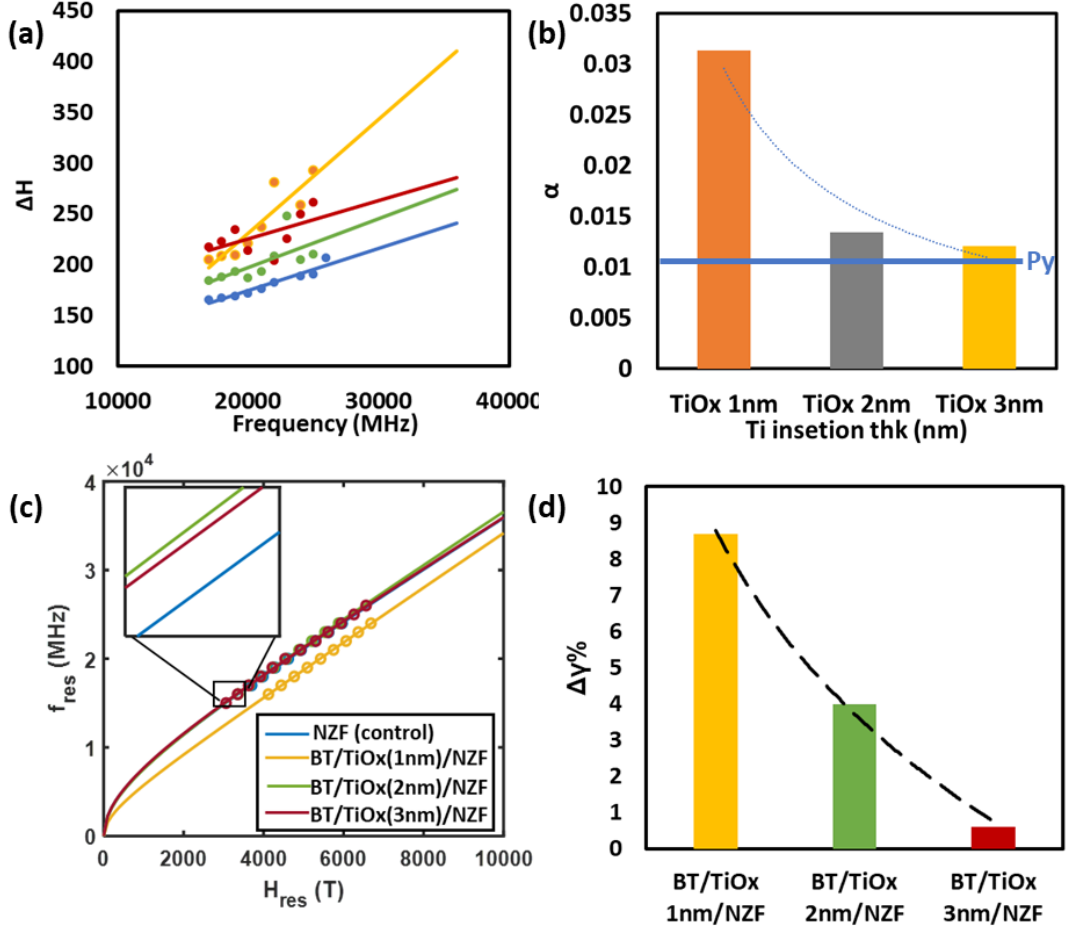

*Figure 6. Results of FMR experiments on $Bi_2Te_3/TiO_x/NZFO$. a) Linewidth ΔH as a function of FMR frequency, b) summary of Gilbert damping and c) FMR frequency as a function of field for NZFO and $Bi_2Te_3/TiO_x/NZFO$. d) Comparison of change in γ compared to the control sample of NZFO because of SOT in the $Bi_2Te_3$ layer.*

### 2.3.3. Interfacial AFM Phase in $Bi_2Te_3/TiO_x/NZFO$ Heterostructures

An AFM interface layer was formed in the $Bi_2Te_3/TiO_x/NZFO$ sample with $z_{TiOx}$ = 1 nm, similar to that discovered in $Bi_2Te_3/Py$ samples. [42] AFM layers form at the interface in $Bi_2Te_3/Py$, and $Bi_2Te_3/TiO_x/Py$ samples as a result of interdiffusion and solid state reaction with the TI surface states [42,43]. In order to understand the effect of interfacial magnetic properties of the $Bi_2Te_3$/$TiO_x$/NZFO heterostructure samples, $m(H)$ loop measurements were performed at 300 K and 6 K similar those shown in Sections 2.1.2 and 2.2. The $m(H)$ loop at 300 K for the $z_{TiOx} = 1$ nm $Bi_2Te_3$/$TiO_x$/NZFO sample was well centered. However, at 6 K the $m(H)$ loop showed a noticeable off-center shift of ~5 Oe signifying exchange bias from the interaction at the AFM-FM interface. The samples with thicker $TiO_x$ (> 2 nm) barrier and the control NZFO sample had a well-centered hysteresis loop at 6 K as well as 300 K, indicating no discernable exchange bias. Table 2 lists values for the exchange bias, $H_{EB}$ and coercive field, $H_C$. This exchange bias is a clear signature of AFM order in the interface emerging in the ferrite heterostructures, again possibly due to solid-state surface chemical reactions promoted by the TSSs of $Bi_2Te_3$. The AFM order in the interface may also arise from PIM in the disordered interface composed of Fe, Bi and Te, similar to V-doped $Sb_2Te_3$/EuS interfaces in ref [47]. These novel magnetic interfacial layers can have topological nontrivial properties as was shown earlier, but the mechanism is still controversial. It is noteworthy that an effective $TiO_x$ barrier, TI/FM as well as TI/MI heterostructures can be grown without changing the crystalline phase of the TI while still maintaining magnetic interactions between MI (or FM) and TSSs. These experimental results are expected to open further exploration of exotic quantum states in interface-engineered TI/MI and TI/FM heterostructures.

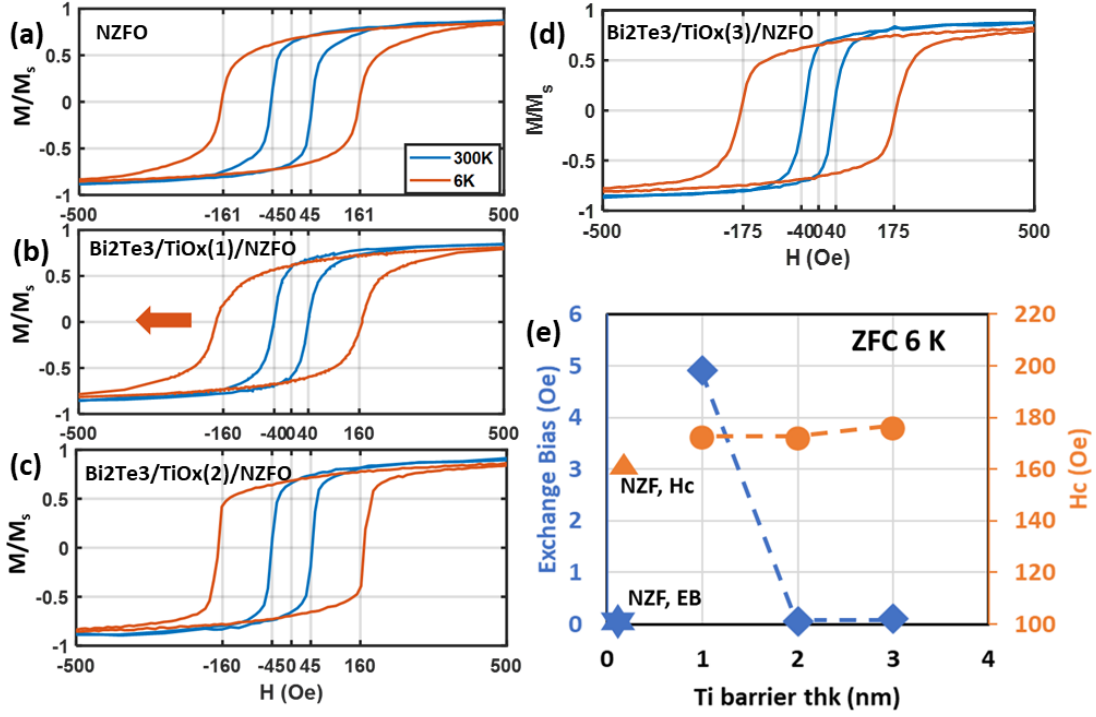

*Figure 7. Magnetic properties of $Bi_2Te_3/TiO_x/NZFO$ for various $TiO_x$ thicknesses. Measured m(H) loops at 300 K and 6 K for: a) NZFO, b) $z_{TiOx}$ = 1 nm, c) $z_{TiOx}$ = 2 nm, and d) $z_{TiOx}$ = 3 nm samples. e) Only the $z_{TiOx}$ = 1 nm sample shows an exchange bias, $H_{EB}$ ≈ 5 Oe arising from formation of AFM interfacial layer at 6 K.*

**Table 2. Properties for $Bi_2Te_3/TiO_x/NZFO$ heterostructures for different $TiO_x$ interlayer thickness. The parameters are described in the text.**

| Material | $z_{TiOx}$ (nm) | $4\pi M_{eff}$ (kOe) | α | $\gamma/2\pi$ (MHz/Oe) | $H_{EB}$ (Oe) | $H_C$ (Oe) |
|---|---|---|---|---|---|---|
| NZFO | - | 6.31 | 0.0118 | 2.81 | 0 | 161.71 |
| $Bi_2Te_3/TiO_x/NZFO$ | 1 | 2.52 | 0.0313 | 3.05 | 4.92 | 171.89 |
| $Bi_2Te_3/TiO_x/NZFO$ | 2 | 5.67 | 0.0134 | 2.92 | 0 | 171.75 |
| $Bi_2Te_3/TiO_x/NZFO$ | 3 | 6.56 | 0.012 | 2.79 | 0 | 175.51 |

## 3. Conclusions

Exposure of TI surface to atmosphere was shown to result in formation of oxides on the surface that are impervious to spin currents and coherent electron hopping. The surface oxidation of TI was shown to magnetically decouple FM grown on them and showed suppression of spin-pumping, interfacial solid-state reactions, and exchange interactions. An interface engineering method with $TiO_x$ barrier was presented. An ultra-thin layer (< 1nm) of $TiO_x$ was shown to protect the TI surface from oxidation without changing the topological properties. This was verified by coupling the TI, $Bi_2Te_3$ with the FM Py, which showed (1) large Gilbert damping enhancement, (2) interface solid-state reaction and (3) exchange bias. With increase in $TiO_x$ thickness beyond 2 nm, the interfacial diffusion can be blocked, while preserving a large spin-pumping effect in $Bi_2Te_3/TiO_x/Py$ heterostructures. Further, low-temperature atmospheric growth of NiZn-ferrite was used as the magnetic insulator in heterostructures. The magnetic properties of NZFO MI on $Bi_2Te_3/TiO_x$ bilayers were studied Similar to the $Bi_2Te_3/TiO_x/Py$ samples with an ultrathin dusting layer of $TiO_x$ (nominally, $z_{TiOx}$ = ~1nm), significant surface reactions were observed for the $z_{TiOx}$ = ~1 nm $Bi_2Te_3/TiO_x/NZFO$ samples, which also showed enhanced Gilbert damping and a small exchange bias at low temperatures (6 K). However, with higher $TiO_x$ barrier thickness, the interface reactions between $Bi_2Te_3$ and NZFO was prevented. The results in this study are expected to open a path towards further exploration of interface engineered TI/FM and TI/MI heterostructures for energy efficient spintronic device applications.

## 4. Experimental Methods

**4.1. Growth of TI/FM and TI/TiO$_x$/FM Heterostructures Using Magnetron Sputtering.** The TI, Bi$_2$Te$_3$ and Py thin films were grown using the same process conditions as the previous work in ref [42]. For studying the effect of surface oxidation, Bi$_2$Te$_3$ was exposed to atmosphere before deposition of FM. Further, Ti thin films were grown using DC magnetron sputtering with 30 W power and 3 mT Ar. The growth rate of Ti thin films was 0.019 nm/s, which were oxidized upon exposure to atmosphere. All the samples were capped with 3 nm of Al on the FM, which was subsequently oxidized to AlO$_x$ on exposure to atmosphere.

**4.2. Low-Temperature Growth of MI, NZFO Using Spin-Spray Process.** Polycrystalline NiZn-ferrite thin films with thickness of 250 nm were deposited on a Si/SiO$_2$ substrate using a home-made 24" spin spray. The spin-spray growth process for NZFO includes a precursor solution with 1.54g/L of FeCl$_2$, 0.04 g/L of ZnCl$_2$, 0.03 g/L of NiCl$_2$ and an oxidizer solution with 0.14 g/L of NaNO$_2$ and 11.48 g/L of CH$_3$COONa. The precursor solution had a pH level of 3-4 and the oxidizer solution had a pH level of 10.8. The process was carried out at a temperature of ~ 100 °C with 500 mL volume of solutions. Detailed discussion of the growth of the NZFO films using the spin-spray process is available in refs [38,39]. XPS measurements of the NZFO samples revealed 77.4 %, 19.3%, 2.0% and 1.2 % average atomic concentrations of O, Fe, Ni and Zn, respectively.

**4.3. Ferromagnetic Resonance.** FMR experiments were performed using field-sweeping at constant frequencies. The FMR signal was detected using a lock-in amplifier for enhanced sensitivity. The RF field for exciting magnetization in the samples was provided by a co-planer waveguide. A pair of Helmholtz coils which produced an AC magnetic field (~500 Hz) was used for the driving field. Control of the experiment and data acquisition was done using NI LabVIEW.

The parameters, FMR linewidth $\Delta H$ and resonance field $H_{res}$ were extracted by fitting a Lorentzian function to the FMR spectra, given by, $\frac{dP}{dH} = K_1 \frac{4\Delta H(H-H_{res})}{[4(H-H_{res})^2+\Delta H^2]} - K_2 \frac{\Delta H^2 - 4(H-H_{res})^2}{[4(H-H_{res})^2+\Delta H^2]}$. Here, $P$ is the absorbed RF power, $K_i (i = 1,2)$ are constants, and $H$ is the applied DC magnetic field.

**4.4. Measurement of *m*(*H*) Hysteresis Loops.** Magnetization *m*(*H*) measurements were obtained using a Quantum Design MPMS XL-7 superconducting quantum interference device (SQUID) magnetometer [42,43]. Hysteresis loop *m*(*H*) measurements were carried out at various temperatures between 6 and 300 K. The ZFC *m*(*T*) measurements were obtained while increasing the temperature in an applied field of 50 Oe. Room temperature *m*(*H*) measurements were also taken using a vibrating sample magnetometer (VSM).

**4.5. TEM and EDS Measurements.** Samples for TEM investigations were prepared by focused ion beam milling (FIB) using a Ga+ ion source [42,43]. Prior to TEM observation an additional cleaning procedure was performed by Ar-ion milling to reduce a surface amorphous layer and residual Ga from the FIB process. The TEM observations were performed using a Talos 200-FX (ThermoFiszher Scientific Inc.) TEM operated at an acceleration voltage of 200 kV. EDS measurements were performed using a ChemiSTEM (ThermoFisher Scientific) and processing of the spectra was performed using Esprit 1.9 (Brucker Inc.) software.

**Acknowledgement**

This work is partially supported by the U.S Army under grant no. W911NF20P0009, the NIH Award UF1NS107694 and by the NSF TANMS ERC Award 1160504. The work of DH and AF was partially supported by the National Science Foundation grant DMR-1905662 and the Air Force Office of Scientific Research award FA9550-20-1-0247. The work of KM was supported by Air Force Research Laboratory under AFRL/NEMO contract: FA8650-19-F-5403 TO3. Studies employing the Titan 60-300 TEM was performed at the Center for Electron Microscopy and Analysis (CEMAS) at The Ohio State University with support through Air Force contract FA8650-18-2-5295. Certain commercial equipments are identified in this paper to foster understanding. Such identification does not imply recommendation or endorsement by Northeastern University and AFRL.

**Conflict of Interest**

The authors declare no conflict of interest.

# Table of Contents

N. Bhattacharjee 1, K. Mahalingam 4, A. Will-Cole 1, Yuyi Wei1, A. Fedorko 2, M. Page 3, M. McConney 3, D. Heiman 2,3, N. X. Sun 1*

**Interface Engineering Enabled Low Temperature Growth of Magnetic Insulator on Topological Insulator**

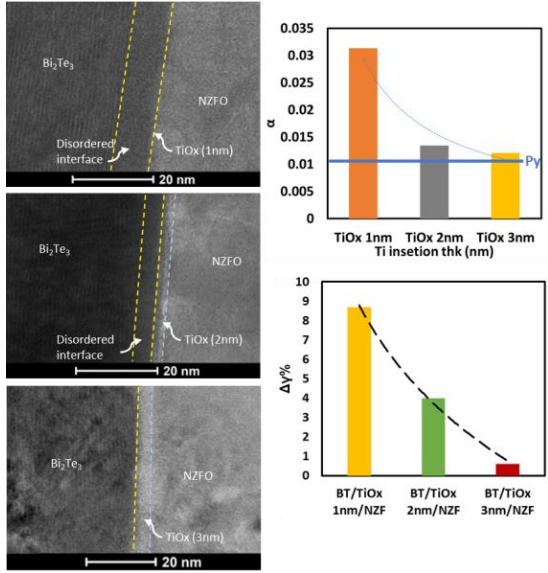

Topological insulators (TIs) are unstable at high temperatures and have highly reactive surfaces. This makes growth of magnetic insulators (MIs) on TIs a significant engineering challenge. In this work, low-temperature growth of MI, NiZn-Ferrite (NZFO) using spin-spray process is demonstrated on TI, $Bi_2Te_3$ capped with thin layers of $TiO_x$. The interface engineered TI/$TiO_x$/MI heterostructures show finite interfacial magnetic interactions without degrading the TI surface.

**Supporting Information: Interface Engineering Enabled Low Temperature Growth of Magnetic Insulator on Topological Insulator**

*Nirjhar Bhattacharjee[1], Krishnamurthy Mahalingam[4], Alexandria Will-Cole[1], Yuyi Wei[1], Adrian Fedorko[2], Cynthia T. Bowers[4], Michael Page[4], Michael McConney[4], Don Heiman[2,3], Nian Xiang Sun[1]\**

1 Northeastern University, Department of Electrical and Computer Engineering, Boston MA 02115, USA

2 Northeastern University, Department of Physics, Boston MA 02115, USA

3 Plasma Science and Fusion Center, MIT, Cambridge, MA 02139, USA

4 Air Force Research Laboratory, Nano-electronic Materials Branch, Wright Patterson Air Force Base, OH 05433, USA
USA

Corresponding author: Nian X. Sun

E-mail: n.sun@northeastern.edu (Corresponding author)

**S1. FMR Spectra for Studying the effect of Surface oxidation of $Bi_2Te_3$**

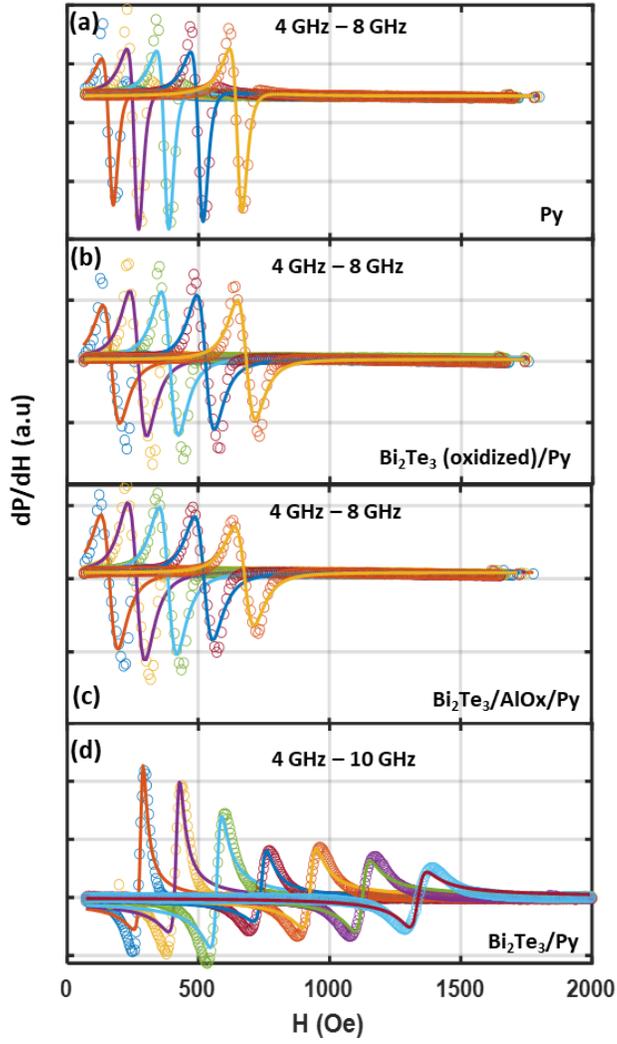

*Figure S1. FMR spectra for Py, $Bi_2Te_3$ (oxidized)/Py, $Bi_2Te_3$/$AlO_x$/Py and $Bi_2Te_3$/Py samples for extracting $\Delta H$, $4\pi M_{eff}$, $H_{res}$ and $\alpha$.*

## S2. FMR Spectra for $Bi_2Te_3$/$TiO_x$/Py Samples

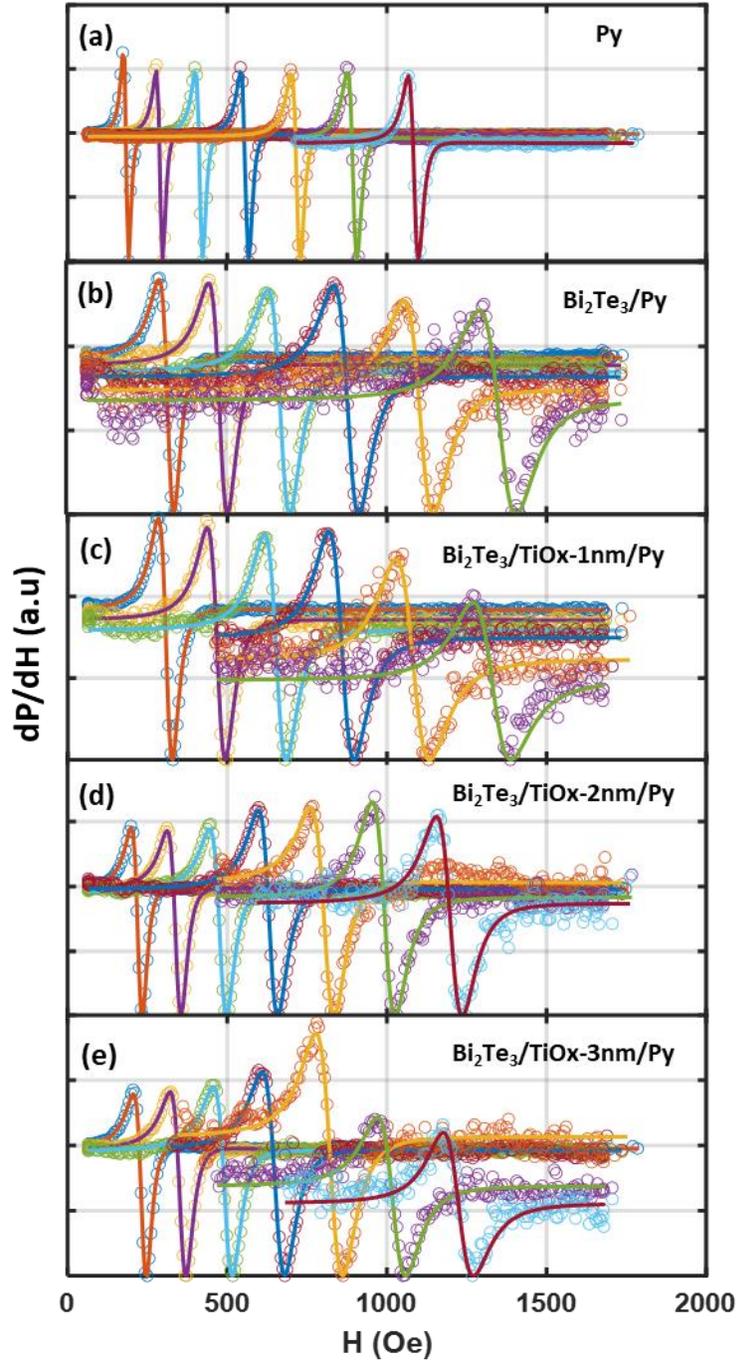

*Figure S2. FMR spectra for Py, $Bi_2Te_3$/Py, $Bi_2Te_3$/TiOx (1,2,3 nm)/Py samples for extracting ΔH, $4\pi M_{eff}$, $H_{res}$ and α.*

## S3. FMR Spectra for $Bi_2Te_3$/TiO$_x$/NZFO Samples

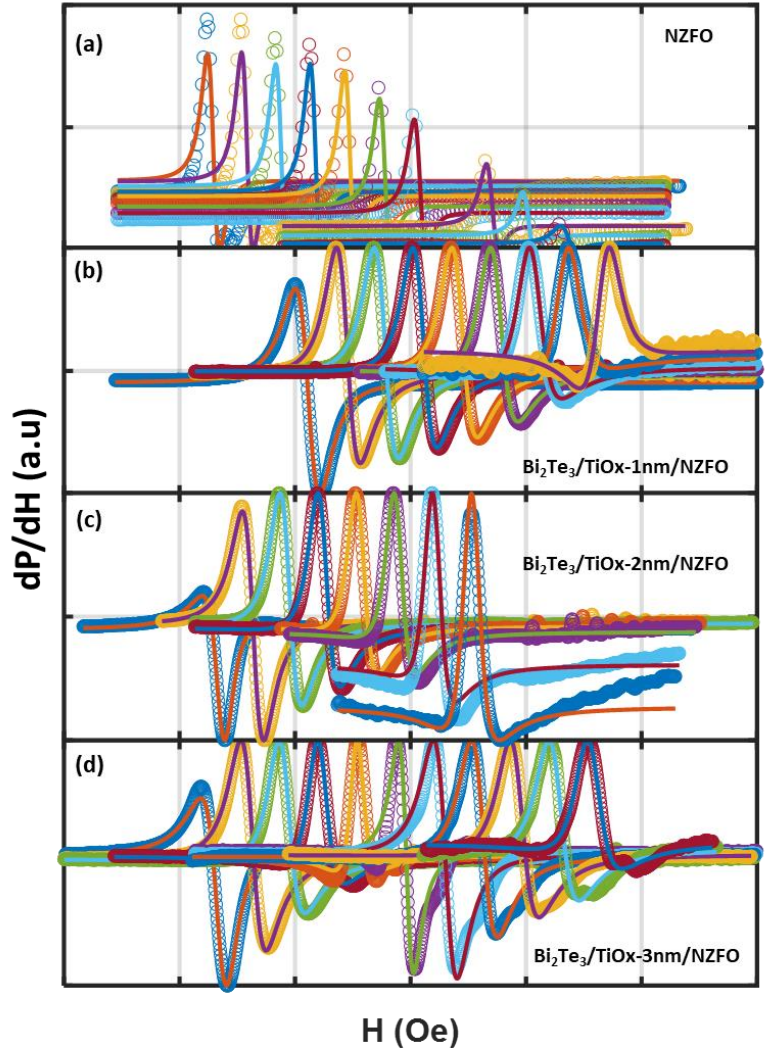

*Figure S3. FMR spectra for NZFO, $Bi_2Te_3$/TiOx (1,2,3 nm)/NZFO samples for extracting $\Delta H$, $4\pi M_{eff}$, $H_{res}$ and $\alpha$.*

**S4. EDS Color Maps and for Studying Interfacial Diffusion**

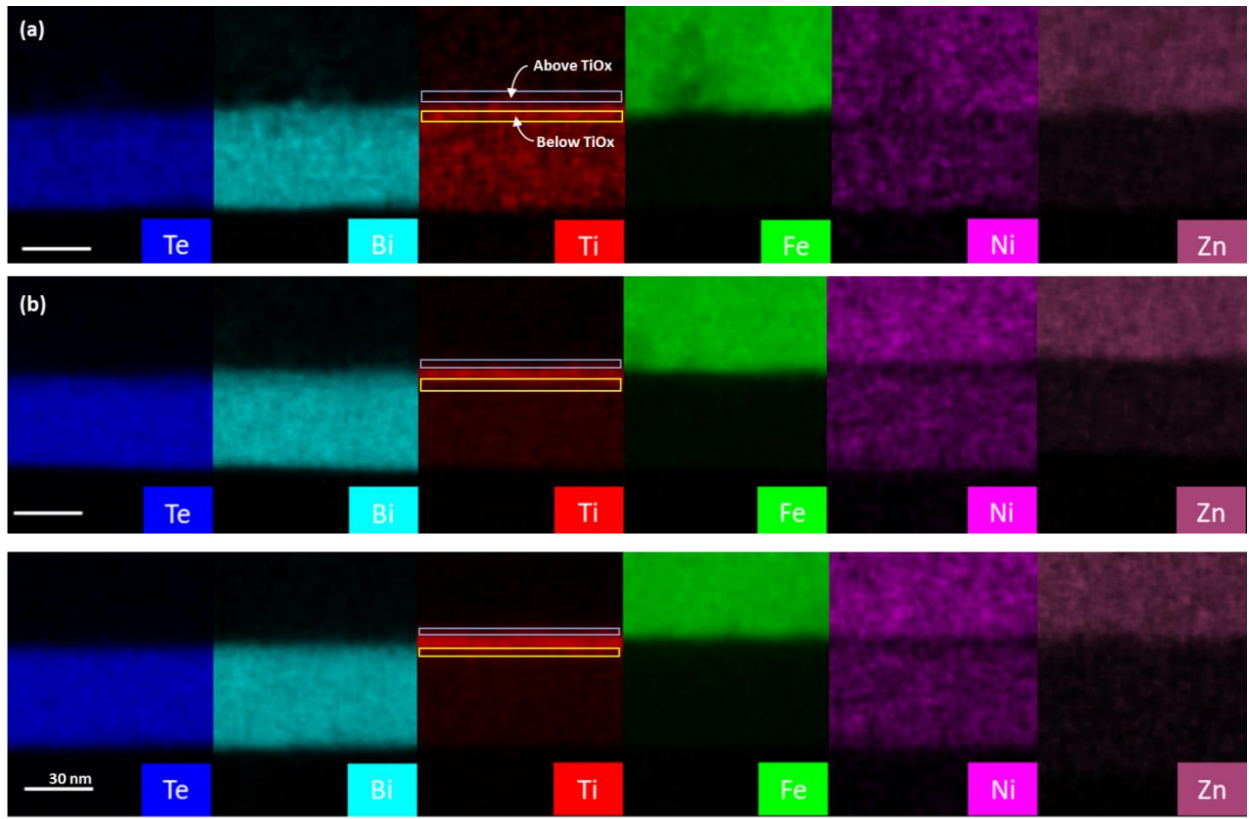

*Figure S4. EDS color maps for studying interfacial diffusion in, a) $Bi_2Te_3$/TiOx (1 nm)/NZFO, b) $Bi_2Te_3$/TiOx (2 nm)/NZFO, and c) $Bi_2Te_3$/TiOx (3 nm)/NZFO samples. The boxes in the Ti color maps denote the above Ti (blue) and below Ti (yellow) regions used for calculation of relative atomic % on Fe, Bi and Te.*

**S5. Repeatability of FMR Results for $Bi_2Te_3$/Py Heterostructures**

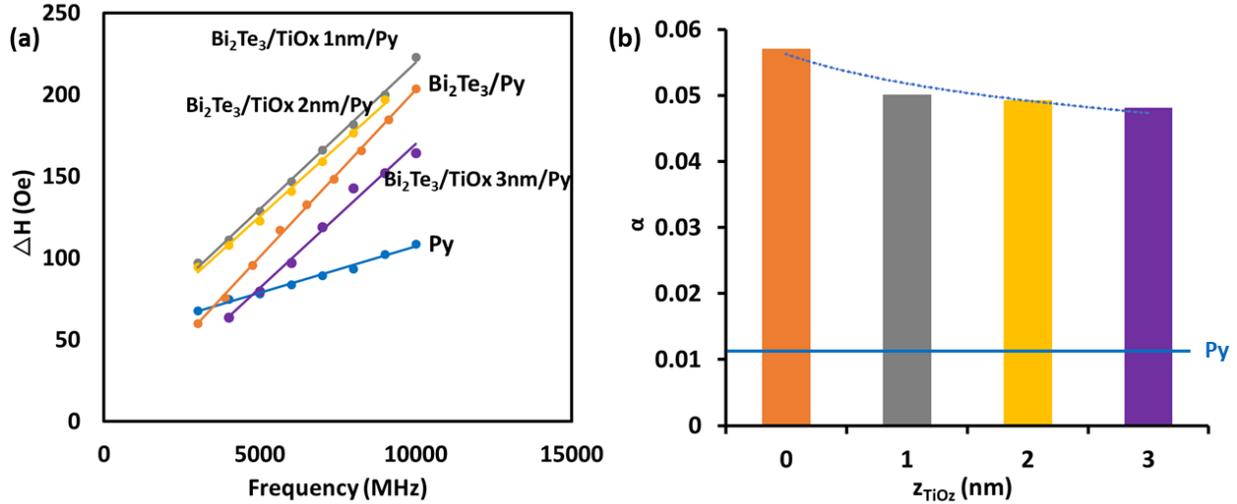

*Figure S5. a) FMR linewidth vs. frequency of $Bi_2Te_3$/Py sample, b) Comparison of Gilbert damping in Py, $Bi_2Te_3$/Py and $Bi_2Te_3$/TiOx/Py samples for varying thickness of barrier TiOx.*

A different set of Py, $Bi_2Te_3$/Py and $Bi_2Te_3$/TiOx/Py samples were grown using the same sputtering process conditions. Similar to the samples in the main text, the TiOx barrier thickness was varied from 1-3 nm. The gilbert damping, $\alpha$ extracted from FMR linewidth and frequency plots, shown in Figure S5 demonstrate a similar trend as the results in the Main Text. The gilbert damping for the control sample of Py was 0.015. However, because of presence of the TI layer and large spin-pumping (also possible SML), the $Bi_2Te_3$/Py and $Bi_2Te_3$/TiO$_x$/Py samples showed large enhancement in α. The $\alpha$ values were 0.0204, 0.0179, 0.0176 and 0.0172 for TiO$_x$ barrier thickness of 0, 1, 2 and 3 nm respectively.